\documentclass[11pt,a4paper]{article}

\usepackage[margin=1in]{geometry}
\usepackage[T1]{fontenc}
\usepackage{booktabs}
\usepackage{amsmath}
\usepackage{array}
\usepackage{xcolor}
\usepackage{graphicx}
\usepackage[hyphens]{url}
\usepackage[hidelinks]{hyperref}
\usepackage{caption}
\usepackage{subcaption}
\usepackage{enumitem}
\usepackage{listings}
\usepackage{fancyvrb}
\usepackage{dirtree}
\usepackage{titlesec}
\titleformat{\section}{\normalfont\large\bfseries}{\thesection}{0.6em}{}
\titleformat{\subsection}{\normalfont\normalsize\bfseries}{\thesubsection}{0.6em}{}

\definecolor{codebg}{RGB}{245,245,245}
\definecolor{codeframe}{RGB}{200,200,200}
\newcommand{\code}[1]{\texttt{\small #1}}

\newcommand{\aiperf}{\textsc{AIPerf}}
\newcommand{\runway}{\textsc{Runway}}

\title{\bfseries Inference Validation for Regulated AI:\\
A Governed Decision Pipeline for LLM Serving}

\author{
  Dev Akre\quad Roman Bobalo\quad Manoj Dobbali\\[2pt]
  \small MasterControl, Inc.\\
  \small\texttt{\{dakre,rbobalo,mdobbali\}@mastercontrol.com}
}
\date{November 9, 2026}

\begin{document}
\maketitle

\begin{abstract}
\noindent
Deploying large language models in regulated industries requires more than picking the fastest
inference engine: it requires \emph{evidence} that a configuration meets contractual and legal
obligations before it serves a single customer request. We describe \runway{}, an internal
validation harness built at MasterControl that builds a governed decision pipeline around
NVIDIA \aiperf{}. The key insight is separating concerns: \aiperf{} provides the trusted measurement,
while \runway{} provides the experiment design that comes before it and the compliance,
cost, and audit decisions that come after. We describe the architecture and the storage model
that makes caching and audit one mechanism. We report three findings from \runway{} sweeps:
TensorRT-LLM and vLLM rankings invert with concurrency; a newer, pricier GPU instance can be
the cheaper one per request; and speculative decoding behaves differently across engines under
load, so engine \emph{configuration} can outweigh engine \emph{choice}.
\end{abstract}

%------------------------------------------------------------------------------
\section{Introduction}
%------------------------------------------------------------------------------

The standard practice for deploying large language models treats inference engine selection as
a one-time decision: benchmark candidates under synthetic load, select the configuration with
the highest throughput, and revisit only when performance degrades. This approach assumes that
faster is better and that the selection criteria are implicit. In regulated industries (pharmaceutical
manufacturing, medical devices, financial compliance), neither assumption holds.

MasterControl operates a quality management platform used by over 1,200 companies in life
sciences and other regulated industries~\cite{mc-about}, under rules such as 21~CFR
Part~11~\cite{cfr11}. Every change to an inference
engine, GPU instance type, or model version is a change to a validated production system.
Regulators do not accept ``the vendor says it's faster'' as justification; they expect
objective, documented evidence that the change was evaluated against defined acceptance
criteria before deployment~\cite{cfr11,fda-csa}.

This paper describes \runway{}, an internal tool that turns the question ``did this
configuration actually meet our requirements?'' into a repeatable, versioned, auditable answer.
The core design decision was to \emph{not} build a benchmarking engine. NVIDIA's \aiperf{}~\cite{aiperf-repo}
already exists, is open-source, and is maintained by the same organization that builds the
GPUs and inference runtimes we deploy on. Instead, \runway{} builds a governed decision
pipeline around an unmodified \aiperf{}: it designs the experiment, invokes \aiperf{} for the measurement, and turns
the result into a release decision it can defend later.

We make two contributions:

\begin{enumerate}[leftmargin=*, label=\textbf{C\arabic*.}]
  \item \textbf{A reference architecture for governed inference validation.} The
    before--measure--after pattern, a storage layout in which caching and audit are one
    mechanism, and campaigns that make a decision question a reviewable artifact
    (\S\ref{sec:architecture}--\S\ref{sec:storage}).
  \item \textbf{Empirical findings on configuration selection.} Engine rankings invert with
    concurrency; the more expensive GPU instance wins on cost per request; and speculative
    decoding separates the engines under load (\S\ref{sec:findings}).
\end{enumerate}

%------------------------------------------------------------------------------
\section{Background: The Measurement Problem}
\label{sec:background}
%------------------------------------------------------------------------------

\subsection{Why LLM Inference Validation Is Different}

Three properties of LLM inference are well known: performance depends on token shape, since
latency grows with output length and time-to-first-token with input length~\cite{nim-params};
outputs are non-deterministic even at temperature zero, so repeated runs are not the same
workload~\cite{he2025}; and speed is not quality, since a quantized model can match average
accuracy yet flip individual answers~\cite{dutta2024}. Quality evaluation is outside the
scope of this paper; a performance gate is necessary but not sufficient. Two further
properties shape \runway{}'s design.

\textbf{Synthetic load is not real load.} Production traffic has a specific
distribution of prompt types, context lengths, and output expectations. Subtle details matter:
engines reuse cached attention state only for an identical prompt
prefix~\cite{vllm-apc,trtllm-kvreuse}, so a system prompt placed first enables prefix
caching, while the same content placed after a per-request field defeats it, changing
time-to-first-token dramatically. Validation must use data that matches production
as closely as possible, or explicitly acknowledge what it does not cover.

\textbf{Results must be traceable.} When asked ``why did you deploy engine X instead of engine
Y,'' the answer must be a retrievable artifact that shows the comparison, the acceptance
criteria, and the pass/fail determination. Part~11 itself requires validated systems and
secure, time-stamped audit trails~\cite{cfr11}.

\subsection{NVIDIA AIPerf}

NVIDIA \aiperf{}~\cite{aiperf-repo}, the successor to GenAI-Perf, is an open-source
benchmark harness for large language model inference. It generates load against an
OpenAI-compatible endpoint, measures time-to-first-token, inter-token latency, and end-to-end
request latency at multiple percentiles, reports throughput in requests and tokens per
second, and samples GPU telemetry alongside~\cite{aiperf-metrics,aiperf-telemetry}. NVIDIA recommends it for
benchmarking generative-AI inference and uses it in its own NIM benchmarking
guide~\cite{nim-bench}.

While \aiperf{} owns the measurement, what a regulated deployment adds around a
measurement is process: deciding which configurations to test, judging whether a result
passes, and keeping the evidence retrievable. That process is what \runway{} supplies.

%------------------------------------------------------------------------------
\section{Architecture: Before, Measure, After}
\label{sec:architecture}
%------------------------------------------------------------------------------

\begin{figure}[t]
\centering
\includegraphics[width=0.82\linewidth]{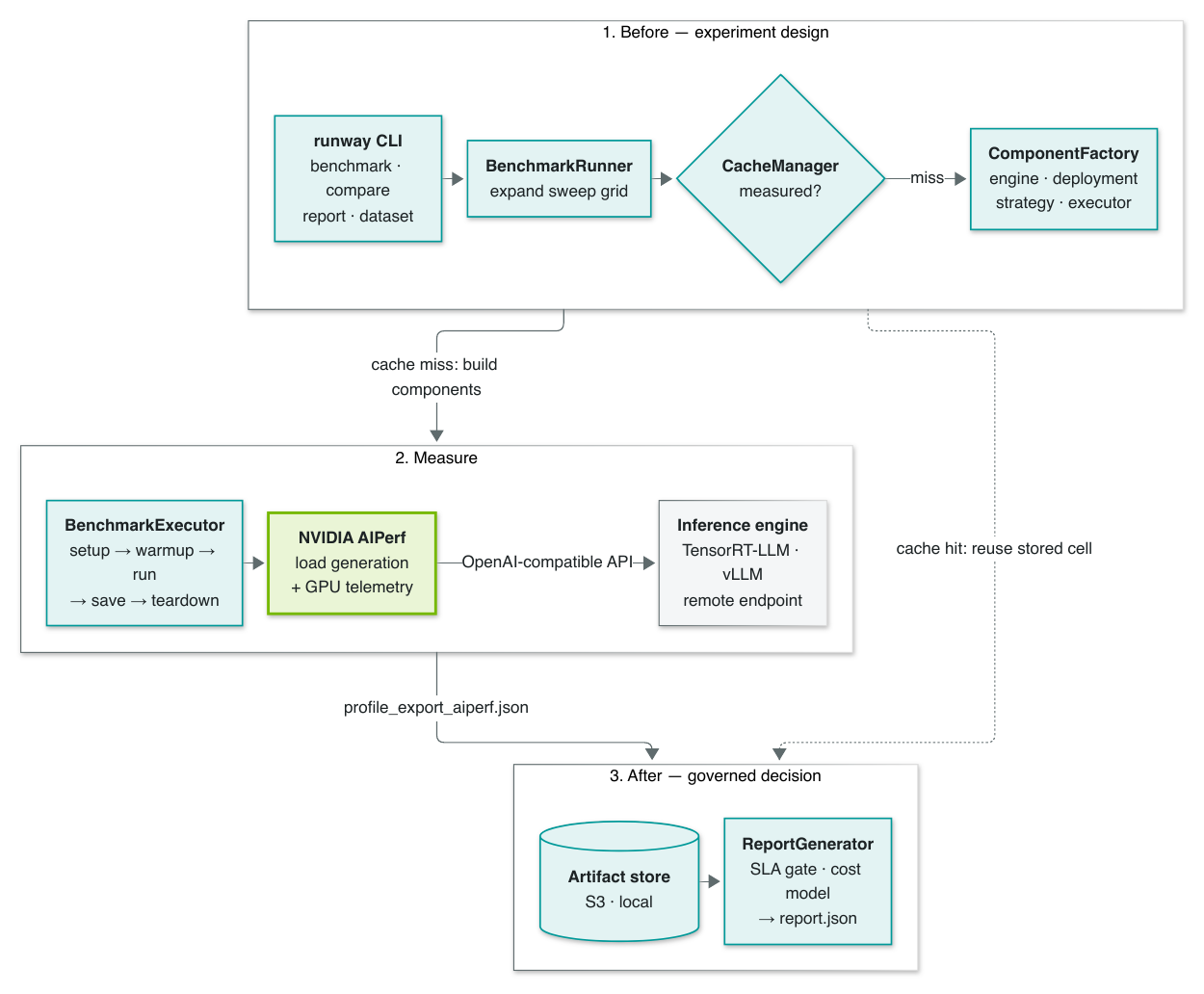}
\caption{\runway{} system overview. Each phase is a separate layer; \aiperf{} is the only
component that touches the engine under test. A cache hit skips measurement entirely.}
\label{fig:architecture}
\end{figure}

\runway{} is a Python CLI organized around three phases (Figure~\ref{fig:architecture}):
\emph{before} the run, sweep expansion and cache lookup; the \emph{measurement}, delegated
entirely to \aiperf{}; and \emph{after} the run, storage, SLA evaluation, and cost modeling.
The separation means the measurement comes from NVIDIA's own tooling (we did not build a load
generator, so we do not have to defend one to auditors), while the decision logic is ours to
change without modifying upstream code.

\subsection{Before: Experiment Design}

A sweep is declared as lists of values for each axis:

\noindent\begin{minipage}{\linewidth}
\begin{lstlisting}[language=bash, caption={Declaring a sweep}]
runway benchmark --model Qwen/Qwen3-8B-FP8 \
  --engines vllm,tensorrt-llm@1.2.1 \
  --instance-type g6e.xlarge --deployment argo \
  --input-tokens 128,512,1024 --output-tokens 128,512,1024,2048 \
  --concurrency 1,5,10,50,100 \
  --gpu-telemetry pynvml --sla sla/production.yaml
\end{lstlisting}
\end{minipage}

\code{BenchmarkRunner} expands this into a grid of cells (2 engines $\times$ 3 input lengths
$\times$ 4 output lengths $\times$ 5 concurrency levels = 120 cells) and asks
\code{CacheManager} which cells were already measured (\S\ref{sec:storage}). Only missing cells
are run.

\subsection{Measure: AIPerf, Unmodified}

\code{BenchmarkExecutor} runs every cell through the same fixed lifecycle (a Template Method):
set up the deployment, start the engine and wait for health, warm up, run, save, tear down. Results
are therefore comparable by construction. For each cell it invokes \aiperf{} as a subprocess;
the command below is the one \runway{} recorded for one of the cells in \S\ref{sec:findings}:

\noindent\begin{minipage}{\linewidth}
\begin{lstlisting}[language=bash, caption={AIPerf invocation for one cell, as recorded}]
aiperf profile --model 'Qwen/Qwen3-8B-FP8' --url 'http://vllm-benchmark:8000' \
  --endpoint-type 'chat' --streaming --concurrency 1 \
  --isl 1024 --osl 1024 --tokenizer 'Qwen/Qwen3-8B-FP8' \
  --gpu-telemetry 'pynvml' --artifact-dir '/workspace/artifacts/result_i1024_o1024_c1'
\end{lstlisting}
\end{minipage}

\runway{} reads \aiperf{}'s \code{profile\_export\_aiperf.json} directly, with no recalculation.
The SLA gate, cost model, and comparison tables all use exactly what \aiperf{} reported.
Engines (TensorRT-LLM, vLLM, or a remote endpoint) and deployment targets are adapters behind
one interface each, so both engines are benchmarked under identical load profiles. When GPU
telemetry is enabled~\cite{aiperf-telemetry}, it also records the device name, which is how
the results below are tied to specific hardware.

\begin{figure}[t]
\centering
\includegraphics[width=0.82\linewidth]{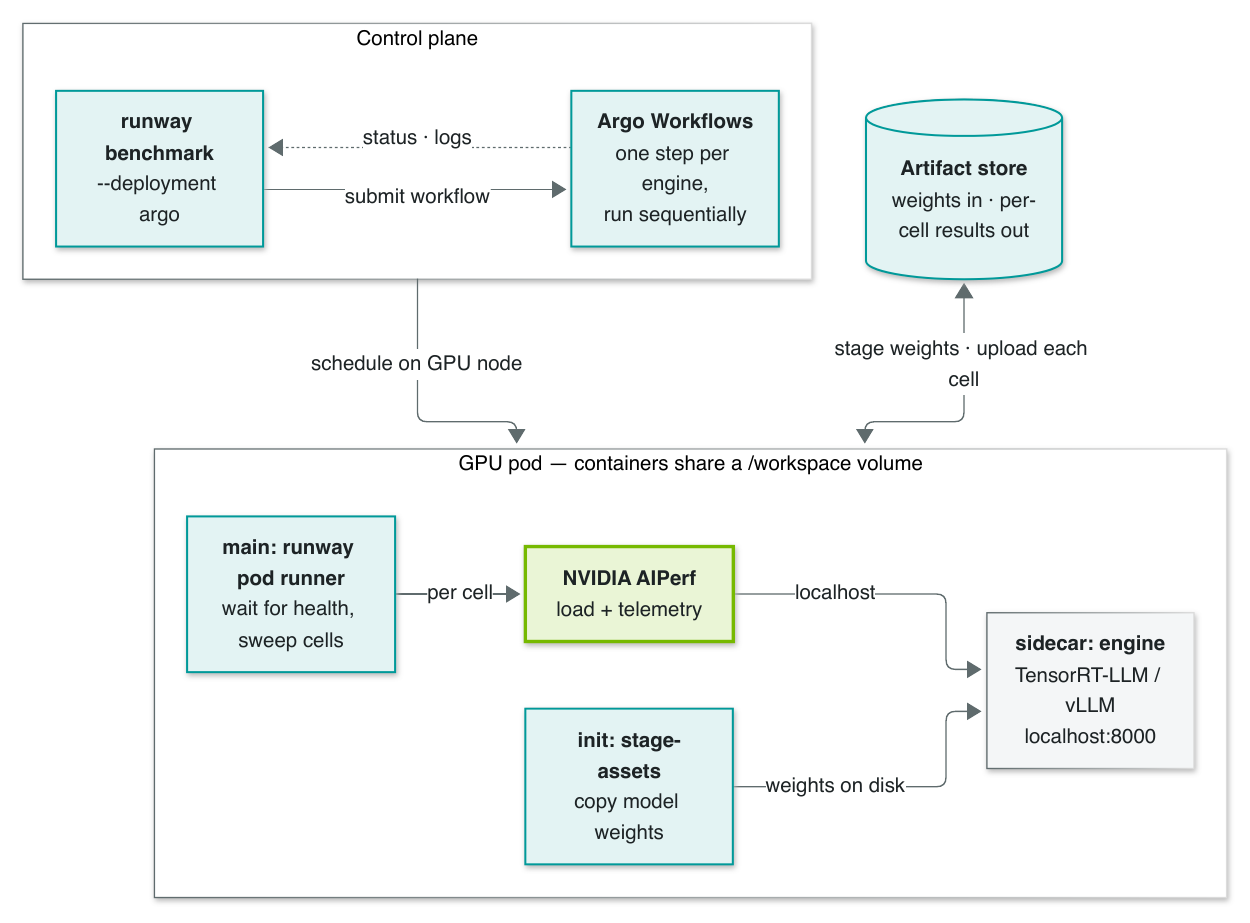}
\caption{Argo Workflows execution. The engine and \aiperf{} share the pod, so load is generated
over \code{localhost}.}
\label{fig:argo}
\end{figure}

On Argo Workflows~\cite{argo} on EKS (Figure~\ref{fig:argo}), an init container stages the
model weights, the engine runs as a sidecar on the pod's shared network, and \runway{} and
\aiperf{} run in the main container against \code{localhost}, with no network hop between the
thing being measured and the thing measuring it. Engines run as sequential steps, so they never
contend for the GPU, and each cell uploads as soon as it finishes, so an interrupted sweep keeps
its completed cells.

\subsection{After: SLA Evaluation and Cost}

Results are evaluated against a declarative, versioned SLA file:

\noindent\begin{minipage}{\linewidth}
\begin{lstlisting}[caption={SLA definition (abridged)}]
name: production
version: 1.2.0
thresholds:
  - metric: request_latency_p95
    operator: "<="
    value: 2.5
    severity: critical
  - metric: ttft_p50
    operator: "<="
    value: 150.0
    severity: warning
\end{lstlisting}
\end{minipage}

Each threshold names an \aiperf{} metric and statistic, a comparison operator, and a severity
(\code{critical}, \code{warning}, or \code{info}). The run rolls up to \code{pass}, \code{partial}
(no critical violation, but a warning violated or a metric missing), or \code{fail}. A cost
model prices the run from live AWS prices. Each run's \code{report.json} records the SLA name,
version, and verdict alongside cost and engine configuration, so ``what thresholds were in
effect when we shipped'' has a stored answer.

For decisions that span runs, \code{runway compare} varies one dimension (model, engine,
instance type, or experiment) and marks the winner per metric. It lists result paths first
and loads only the matching cells, so a comparison over hundreds of cells needs no GPU and
takes seconds.

%------------------------------------------------------------------------------
\section{Configuration and Campaigns}
\label{sec:config}
%------------------------------------------------------------------------------

Engine settings live in one YAML file per engine and model, passed to \code{trtllm-serve} or
\code{vllm serve} unchanged. A \emph{model spec} bundles a recurring sweep into one file and can
pass \aiperf{} options straight through (for example, token-length spread instead of fixed
lengths). A single \code{versions.yaml} pins the engine images and the \aiperf{} release; the
version is part of every result path, so results from different versions never mix. Workloads
are either synthetic token shapes generated by \aiperf{} or an uploaded prompt set, single- or
multi-turn; the dataset name is the first path segment, and reports refer to datasets only by
name, so a report can be shared without its prompts.

A single sweep answers ``how fast?''. A \emph{campaign} answers a decision question such as ``does
speculative decoding pay off for this workload?'' and is kept as a directory of reviewable
text:

\begin{itemize}[leftmargin=*]
  \item \code{manifest.yaml}: the contract, holding the question, model, engine versions, sweep
    settings, datasets with \emph{measured} input/output lengths, and the comparison structure.
    A \emph{leg} fixes one hardware, dataset, and cache regime; each \emph{arm} is one engine
    configuration within it, with its own provenance.
  \item \code{configs/}: engine configurations exactly as submitted.
  \item \code{run.sh}: how to reproduce; \code{FINDINGS.md}: numbers, recommendation,
    caveats.
\end{itemize}

A validator rejects any manifest in which two arms would write to the same result
prefix, a collision that would otherwise silently overwrite one arm with another. The findings
in \S\ref{sec:mtp} come from two such campaigns.

%------------------------------------------------------------------------------
\section{Result Storage and Caching}
\label{sec:storage}
%------------------------------------------------------------------------------

Every cell's result lands at a deterministic path (Figure~\ref{fig:storage}). The path answers
\emph{what} was measured: dataset, instance, model, engine, version, optional experiment name,
and the cell's input length, output length, and concurrency. A second file next to
\aiperf{}'s output, \code{benchmark\_config.json}, records \emph{how}: streaming, extra \aiperf{} arguments,
duration, dataset format, warm-up, and the engine configuration.

% Fig. 2: artifact store layout (storage/paths.py). Root is symbolic on purpose;
% do not replace <bucket> with a real bucket or prefix.
\begin{figure}[t]
\centering
\begin{minipage}{0.92\linewidth}
\small
\dirtree{%
.1 \texttt{s3://<bucket>/runway/}.
.2 \texttt{benchmarks/<dataset>/<instance>/<model>/<engine>/<version>/[<experiment>/]}.
.3 \texttt{result\_i<I>\_o<O>\_c<C>/} \dotfill{} \textit{one sweep cell}.
.4 \texttt{profile\_export\_aiperf.json} \dotfill{} \textit{AIPerf output, unmodified}.
.4 \texttt{benchmark\_config.json} \dotfill{} \textit{cache fingerprint}.
.2 \texttt{reports/<instance>/<org>/<model>/<engine>/<timestamp>/report.json} \dotfill{} \textit{SLA + cost verdict}.
.2 \texttt{datasets/<name>/dataset.txt} \dotfill{} \textit{uploaded prompt sets}.
.2 \texttt{campaigns/<id>/results/} \dotfill{} \textit{normalized campaign snapshots}.
}
\end{minipage}
\caption{Artifact store layout. The path encodes \emph{what} was measured; the fingerprint
file records \emph{how}. Synthetic sweeps use \texttt{synthetic} as the dataset segment.}
\label{fig:storage}
\end{figure}

\begin{figure}[t]
\centering
\includegraphics[width=0.78\linewidth]{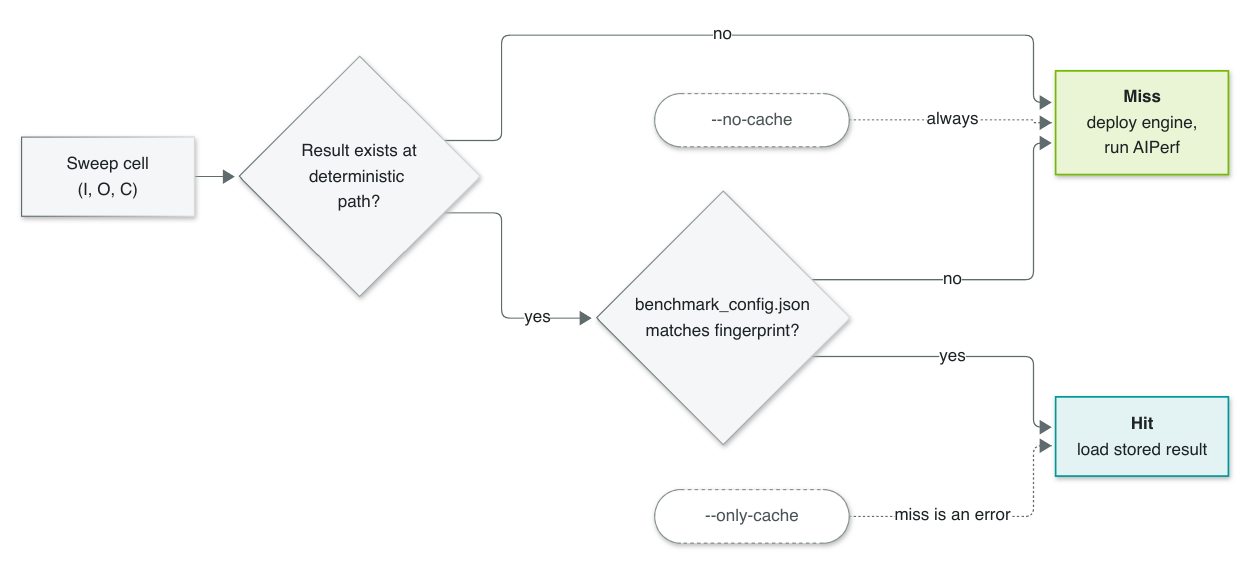}
\caption{Per-cell cache decision. \code{-{}-only-cache} turns any miss into an error, which lets a
report be regenerated on a machine with no GPU.}
\label{fig:cache}
\end{figure}

The same layout makes caching and audit one mechanism. Before deploying anything,
\code{CacheManager} checks each cell (Figure~\ref{fig:cache}): the result must exist \emph{and}
its fingerprint must match the requested run. The fingerprint compares the settings that
change what \aiperf{} measures (\code{streaming}, \code{aiperf\_extra\_args},
\code{benchmark\_duration}, \code{custom\_dataset\_type}). The engine configuration is
recorded but not compared; a run with a different engine configuration takes an
\code{-{}-experiment} name, which adds a path segment and so a fresh cache slot. The rule
generalizes: anything that changes a result must change either its path or its fingerprint.

A full hit reproduces a report without a GPU. The cost is that a repeated run at the same
configuration returns the stored numbers rather than a new measurement, so each cell has
exactly one independent measurement unless \code{-{}-no-cache} is used (\S\ref{sec:lessons}).

%------------------------------------------------------------------------------
\section{Findings}
\label{sec:findings}
%------------------------------------------------------------------------------

We report three findings from \runway{} sweeps. All numbers come from stored cells; each cell
is a single measurement (\S\ref{sec:storage}), so small differences should be read as
indicative rather than significant.

\subsection{Engine Rankings Invert with Concurrency}

\textbf{Setup.} vLLM 0.19.1 and TensorRT-LLM 1.2.1 at default settings on Qwen3-8B-FP8~\cite{qwen3},
60 matched cells: 12 token shapes (input $\in \{128, 512, 1024\}$, output
$\in \{128, 512, 1024, 2048\}$) $\times$ 5 concurrency levels (1, 5, 10, 50, 100), on
\code{g6e.xlarge} (NVIDIA L40S, 48\,GB~\cite{aws-g6e}), \aiperf{} 0.7.0.

\begin{table}[h]
\centering
\small
\begin{tabular}{lcc}
\toprule
\textbf{Concurrency} & \textbf{Throughput Winner} & \textbf{p95 Latency Winner} \\
\midrule
1   & TensorRT-LLM (+35\%) & TensorRT-LLM (12/12) \\
5   & TensorRT-LLM (+13\%) & TensorRT-LLM (12/12) \\
10  & TensorRT-LLM (+10\%) & TensorRT-LLM (12/12) \\
50  & vLLM (+3\%)          & TensorRT-LLM (7/12)  \\
100 & vLLM (+19\%)         & vLLM (12/12)         \\
\bottomrule
\end{tabular}
\caption{Engine comparison across concurrency levels. Throughput advantage is the winner's
mean throughput ratio over the 12 token shapes; latency winner counts wins out of 12 shapes.}
\label{tab:engine-comparison}
\end{table}

The crossover occurs between concurrency 10 and 50. Both engines implement a paged KV cache
and continuous (in-flight) batching~\cite{vllm2023,trtllm-repo}, so we read it as a default
scheduler setting rather than an architectural difference, though these sweeps do not isolate
the cause. At low concurrency, TensorRT-LLM's per-token speed dominates. At high concurrency,
its default scheduler policy, \code{GUARANTEED\_NO\_EVICT}, uses the KV cache conservatively,
guaranteeing that a started request is never paused, and so admits fewer requests at
once~\cite{trtllm-sched}. \S\ref{sec:mtp}
shows how much a different policy changes the picture.

A single-concurrency benchmark can therefore pick either engine depending on the level chosen;
engine selection should follow the deployment's concurrency profile, and comparisons
should test several levels, as NVIDIA's NIM benchmarking guidance already
recommends~\cite{nim-params}.

\subsection{The Pricier Instance Is Cheaper per Request}

\textbf{Setup.} The same model on vLLM 0.19.1, 80 matched cells (16 token shapes, input up to
4,096, $\times$ 5 concurrency levels) on \code{g6e.xlarge} (NVIDIA L40S) and
\code{g7e.2xlarge} (NVIDIA RTX PRO 6000 Blackwell Server Edition, 96\,GB~\cite{aws-g7e}). The
cost model priced them at \$1.861 and \$3.363 per hour (us-west-2, Linux on
demand, as of 2026-09~\cite{aws-ec2-od}), a 1.81$\times$ premium for g7e.

\begin{table}[h]
\centering
\small
\begin{tabular}{lccc}
\toprule
\textbf{Concurrency} & \textbf{Throughput} & \textbf{p50 Latency} & \textbf{Cost per Request} \\
\midrule
1   & 2.13$\times$ & 0.47$\times$ & 0.85$\times$ \\
5   & 1.82$\times$ & 0.55$\times$ & 0.99$\times$ \\
10  & 1.85$\times$ & 0.54$\times$ & 0.98$\times$ \\
50  & 1.94$\times$ & 0.52$\times$ & 0.93$\times$ \\
100 & 2.04$\times$ & 0.49$\times$ & 0.88$\times$ \\
\bottomrule
\end{tabular}
\caption{\code{g7e.2xlarge} relative to \code{g6e.xlarge}, median over 16 token shapes. Cost per
request is the price ratio divided by the throughput ratio.}
\label{tab:hardware}
\end{table}

Blackwell roughly doubles throughput and halves latency at every concurrency, in line with
AWS's own figure of up to 2.3$\times$ the inference performance of G6e~\cite{aws-g7e}, which more
than pays for its price: the median cost per request is 1--15\% lower (Table~\ref{tab:hardware}).
This is the ``after'' phase at work. A cost-per-hour view picks the cheaper instance, while
the measured cost per request picks the other one. One cell (input and output 128,
concurrency 1) inverts, with g7e slower; with a single measurement per cell we cannot yet tell
whether that is real.

\subsection{Speculative Decoding: Configuration Can Outweigh Engine Choice}
\label{sec:mtp}

\textbf{Setup.} Two campaigns (\S\ref{sec:config}) on Qwen3.5-9B~\cite{qwen35-9b} with one
NVIDIA H100 (\code{p5.4xlarge}, 80\,GB~\cite{aws-p5}): an internal entity-extraction prompt set
with a shared prefix (about 3,300 input and 200 output tokens), 60-second cells, cache disabled.
TensorRT-LLM 1.3.0rc23 (a pre-release; the pinned 1.2.1 did not support Multi-Token Prediction
for this model) against vLLM 0.19.1, both tuned. Each ran with and without Multi-Token
Prediction (MTP)~\cite{gloeckle2024}, a form of speculative decoding~\cite{leviathan2023}
that uses the model's own MTP head to draft tokens~\cite{vllm-mtp,trtllm-specdec}, at five draft
tokens. The model card suggests two draft tokens for vLLM~\cite{qwen35-9b}, so the longer draft
may contribute to vLLM's result under load.

\begin{table}[h]
\centering
\small
\begin{tabular}{lcc}
\toprule
\textbf{Concurrency} & \textbf{vLLM: MTP / baseline} & \textbf{TensorRT-LLM: MTP / baseline} \\
\midrule
1  & 2.16$\times$ & 2.57$\times$ \\
16 & 1.11$\times$ & 1.69$\times$ \\
32 & 0.92$\times$ & 1.32$\times$ \\
64 & 0.80$\times$ & 1.45$\times$ \\
\bottomrule
\end{tabular}
\caption{Throughput with MTP relative to the same engine without it.}
\label{tab:mtp}
\end{table}

MTP cuts TensorRT-LLM's median inter-token latency at concurrency 1 from 6.65 to 2.18\,ms, a
3$\times$ speed-up per token. Speculation pays off most when small batches leave the GPU
underused~\cite{trtllm-specdec}, so both gains shrink with load, but the engines diverge in how
far (Table~\ref{tab:mtp}). vLLM's MTP turns into a net cost above roughly concurrency 16, while
TensorRT-LLM's stays positive at every level. So each engine's best configuration differs in
kind: vLLM performs best without MTP, TensorRT-LLM with it. At concurrency 64, vLLM's best still
leads, 20.9 against 19.3\,req/s. It gets there on defaults, whereas TensorRT-LLM needed a
non-default scheduler policy (\code{MAX\_UTILIZATION} instead of
\code{GUARANTEED\_NO\_EVICT}~\cite{trtllm-sched}). The campaign notes attribute a 2.55$\times$
throughput gain to that one setting. The lesson for the comparison in
Table~\ref{tab:engine-comparison} is that it holds \emph{at default settings}: a single engine
setting can move throughput by more than the gap between the engines.

%------------------------------------------------------------------------------
\section{Discussion and Limitations}
\label{sec:lessons}
%------------------------------------------------------------------------------

\textbf{Cost of a custom load generator.} Before adopting \aiperf{}, we prototyped an in-house
benchmark harness. It took three engineer-weeks to reach feature parity with \aiperf{}, and
its correctness would still have needed independent justification during validation. Adopting
\aiperf{} unmodified moved that effort from measurement to the decision pipeline around it.

\textbf{Caching versus variance.} Fingerprint caching makes reports reproducible and avoids
re-measuring known cells, but it also means each cell holds exactly one measurement, so none of
the results in \S\ref{sec:findings} carry confidence intervals. A replication option that
deliberately bypasses the cache (\code{-{}-replicate=N}) is planned and was not available for
these sweeps.

\textbf{Synthetic workloads.} The findings in Tables~\ref{tab:engine-comparison}
and~\ref{tab:hardware} use generated token shapes, not replayed production traffic, and the
production token distribution has not yet been measured. They establish relative engine and
instance behavior, not performance on any specific MasterControl workload.

%------------------------------------------------------------------------------
\section{Conclusion}
\label{sec:conclusion}
%------------------------------------------------------------------------------

Regulated AI deployments require evidence, not intuition. We described \runway{}, a
validation harness that builds a governed decision pipeline around NVIDIA \aiperf{} by designing
experiments before the measurement, applying declarative SLA and cost gates after, and storing
every result at a path that says what it measured. The key design insight is that measurement
and decision are different problems with different owners: \aiperf{} provides the trusted
measurement, \runway{} provides the process around it.

The findings make the same point from the data. Engine rankings invert with concurrency, the
more expensive instance can be the cheaper one per request, and one engine setting can outweigh
the choice of engine. Each of these is invisible to a single-point benchmark and plain in a
governed sweep.

\runway{} is an internal tool at MasterControl and is not currently open-source. We welcome
inquiries from practitioners building similar harnesses in regulated settings.

%------------------------------------------------------------------------------
% References
%------------------------------------------------------------------------------
\bibliographystyle{IEEEtran}
\bibliography{references}

\begin{thebibliography}{10}
\providecommand{\url}[1]{#1}
\csname url@samestyle\endcsname
\providecommand{\newblock}{\relax}
\providecommand{\bibinfo}[2]{#2}
\providecommand{\BIBentrySTDinterwordspacing}{\spaceskip=0pt\relax}
\providecommand{\BIBentryALTinterwordstretchfactor}{4}
\providecommand{\BIBentryALTinterwordspacing}{\spaceskip=\fontdimen2\font plus
\BIBentryALTinterwordstretchfactor\fontdimen3\font minus
  \fontdimen4\font\relax}
\providecommand{\BIBforeignlanguage}[2]{{%
\expandafter\ifx\csname l@#1\endcsname\relax
\typeout{** WARNING: IEEEtran.bst: No hyphenation pattern has been}%
\typeout{** loaded for the language `#1'. Using the pattern for}%
\typeout{** the default language instead.}%
\else
\language=\csname l@#1\endcsname
\fi
#2}}
\providecommand{\BIBdecl}{\relax}
\BIBdecl

\bibitem{mc-about}
\BIBentryALTinterwordspacing
{MasterControl}, ``{MasterControl Strengthens Executive Leadership Team with
  Two Key Appointments},'' Press release, May 2026, accessed: 2026-09.
  [Online]. Available:
  \url{https://www.mastercontrol.com/news/mastercontrol-strengthens-executive-team-two-key-appointments}
\BIBentrySTDinterwordspacing

\bibitem{cfr11}
\BIBentryALTinterwordspacing
{U.S. Government}, ``{Electronic Records; Electronic Signatures},'' Code of
  Federal Regulations, Title 21, Part 11, via eCFR, accessed: 2026-09.
  [Online]. Available:
  \url{https://www.ecfr.gov/current/title-21/chapter-I/subchapter-A/part-11}
\BIBentrySTDinterwordspacing

\bibitem{fda-csa}
\BIBentryALTinterwordspacing
{U.S. Food and Drug Administration}, ``{Computer Software Assurance for
  Production and Quality Management System Software: Guidance for Industry and
  Food and Drug Administration Staff},'' Feb. 2026, accessed: 2026-09.
  [Online]. Available:
  \url{https://www.fda.gov/regulatory-information/search-fda-guidance-documents/computer-software-assurance-production-and-quality-management-system-software}
\BIBentrySTDinterwordspacing

\bibitem{aiperf-repo}
\BIBentryALTinterwordspacing
{NVIDIA}, ``{AIPerf},'' GitHub repository, accessed: 2026-09. [Online].
  Available: \url{https://github.com/ai-dynamo/aiperf}
\BIBentrySTDinterwordspacing

\bibitem{nim-params}
\BIBentryALTinterwordspacing
------, ``{NVIDIA NIM LLMs Benchmarking Guide: Parameters and Best
  Practices},'' accessed: 2026-10. [Online]. Available:
  \url{https://docs.nvidia.com/nim/benchmarking/llm/parameters}
\BIBentrySTDinterwordspacing

\bibitem{he2025}
\BIBentryALTinterwordspacing
H.~He and {Thinking Machines Lab}, ``Defeating nondeterminism in {LLM}
  inference,'' Thinking Machines Lab: Connectionism, Sep. 2025, accessed:
  2026-09. [Online]. Available:
  \url{https://thinkingmachines.ai/blog/defeating-nondeterminism-in-llm-inference/}
\BIBentrySTDinterwordspacing

\bibitem{dutta2024}
A.~Dutta, S.~Krishnan, N.~Kwatra, and R.~Ramjee, ``Accuracy is not all you
  need,'' in \emph{Advances in Neural Information Processing Systems
  (NeurIPS)}, 2024.

\bibitem{vllm-apc}
\BIBentryALTinterwordspacing
{vLLM Project}, ``{Automatic Prefix Caching},'' accessed: 2026-09. [Online].
  Available:
  \url{https://docs.vllm.ai/en/latest/features/automatic_prefix_caching/}
\BIBentrySTDinterwordspacing

\bibitem{trtllm-kvreuse}
\BIBentryALTinterwordspacing
{NVIDIA}, ``{KV Cache System},'' TensorRT-LLM documentation, accessed: 2026-10.
  [Online]. Available:
  \url{https://nvidia.github.io/TensorRT-LLM/features/kvcache.html}
\BIBentrySTDinterwordspacing

\bibitem{aiperf-metrics}
\BIBentryALTinterwordspacing
------, ``{AIPerf Metrics Reference},'' accessed: 2026-09. [Online]. Available:
  \url{https://github.com/ai-dynamo/aiperf/blob/main/docs/metrics-reference.md}
\BIBentrySTDinterwordspacing

\bibitem{aiperf-telemetry}
\BIBentryALTinterwordspacing
------, ``{AIPerf GPU Telemetry Tutorial},'' accessed: 2026-09. [Online].
  Available:
  \url{https://github.com/ai-dynamo/aiperf/blob/main/docs/tutorials/gpu-telemetry.md}
\BIBentrySTDinterwordspacing

\bibitem{nim-bench}
\BIBentryALTinterwordspacing
------, ``{NVIDIA NIM LLMs Benchmarking Guide: Overview},'' accessed: 2026-10.
  [Online]. Available:
  \url{https://docs.nvidia.com/nim/benchmarking/llm/overview}
\BIBentrySTDinterwordspacing

\bibitem{argo}
\BIBentryALTinterwordspacing
{Argo Project}, ``{Argo Workflows: The Workflow Engine for Kubernetes},''
  accessed: 2026-09. [Online]. Available:
  \url{https://argo-workflows.readthedocs.io/en/latest/}
\BIBentrySTDinterwordspacing

\bibitem{qwen3}
A.~Yang \emph{et~al.}, ``Qwen3 technical report,'' arXiv preprint
  arXiv:2505.09388, 2025.

\bibitem{aws-g6e}
\BIBentryALTinterwordspacing
{Amazon Web Services}, ``{Amazon EC2 G6e Instances},'' accessed: 2026-09.
  [Online]. Available: \url{https://aws.amazon.com/ec2/instance-types/g6e/}
\BIBentrySTDinterwordspacing

\bibitem{vllm2023}
W.~Kwon \emph{et~al.}, ``Efficient memory management for large language model
  serving with {PagedAttention},'' in \emph{Proc. SOSP}, 2023.

\bibitem{trtllm-repo}
\BIBentryALTinterwordspacing
{NVIDIA}, ``{Paged Attention, IFB, and Request Scheduling},'' TensorRT-LLM
  documentation, accessed: 2026-10. [Online]. Available:
  \url{https://nvidia.github.io/TensorRT-LLM/features/paged-attention-ifb-scheduler.html}
\BIBentrySTDinterwordspacing

\bibitem{trtllm-sched}
\BIBentryALTinterwordspacing
------, ``{Useful Runtime Options: Capacity Scheduler Policy},'' TensorRT-LLM
  Performance Tuning Guide, accessed: 2026-09. [Online]. Available:
  \url{https://nvidia.github.io/TensorRT-LLM/performance/performance-tuning-guide/useful-runtime-flags.html}
\BIBentrySTDinterwordspacing

\bibitem{aws-g7e}
\BIBentryALTinterwordspacing
{Amazon Web Services}, ``{Amazon EC2 G7e Instances},'' accessed: 2026-09.
  [Online]. Available: \url{https://aws.amazon.com/ec2/instance-types/g7e/}
\BIBentrySTDinterwordspacing

\bibitem{aws-ec2-od}
\BIBentryALTinterwordspacing
------, ``{Amazon EC2 On-Demand Pricing},'' accessed: 2026-09. [Online].
  Available: \url{https://aws.amazon.com/ec2/pricing/on-demand/}
\BIBentrySTDinterwordspacing

\bibitem{qwen35-9b}
\BIBentryALTinterwordspacing
{Qwen Team}, ``{Qwen3.5-9B Model Card},'' Hugging Face, accessed: 2026-09.
  [Online]. Available: \url{https://huggingface.co/Qwen/Qwen3.5-9B}
\BIBentrySTDinterwordspacing

\bibitem{aws-p5}
\BIBentryALTinterwordspacing
{Amazon Web Services}, ``{Amazon EC2 P5 Instances},'' accessed: 2026-09.
  [Online]. Available: \url{https://aws.amazon.com/ec2/instance-types/p5/}
\BIBentrySTDinterwordspacing

\bibitem{gloeckle2024}
F.~Gloeckle \emph{et~al.}, ``Better \& faster large language models via
  multi-token prediction,'' in \emph{Proc. 41st International Conference on
  Machine Learning (ICML)}, ser. PMLR, vol. 235, 2024, pp. 15\,706--15\,734.

\bibitem{leviathan2023}
Y.~Leviathan, M.~Kalman, and Y.~Matias, ``Fast inference from transformers via
  speculative decoding,'' in \emph{Proc. ICML}, 2023.

\bibitem{vllm-mtp}
\BIBentryALTinterwordspacing
{vLLM Project}, ``{MTP (Multi-Token Prediction)},'' accessed: 2026-09.
  [Online]. Available:
  \url{https://docs.vllm.ai/en/latest/features/speculative_decoding/mtp/}
\BIBentrySTDinterwordspacing

\bibitem{trtllm-specdec}
\BIBentryALTinterwordspacing
{NVIDIA}, ``{Speculative Decoding},'' TensorRT-LLM documentation, accessed:
  2026-09. [Online]. Available:
  \url{https://nvidia.github.io/TensorRT-LLM/features/speculative-decoding.html}
\BIBentrySTDinterwordspacing

\end{thebibliography}

\end{document}